\documentclass[aps,prb,twocolumn,nofootinbib,superscriptaddress]{revtex4-1}
\usepackage[a4paper,top=1.40cm,bottom=1.40cm,left=1.75cm,right=1.75cm]{geometry}
\usepackage{mathptmx}
\usepackage{graphicx}
\DeclareGraphicsExtensions{.eps, .pdf, .jpg, .png}
\usepackage[pdftex,colorlinks=true,linkcolor=blue,citecolor=blue]{hyperref}

\usepackage{amsmath}
\usepackage{nicefrac}
\usepackage{float}
\usepackage[T1]{fontenc}
\newcommand\thefont{\expandafter\string\the\font}
\usepackage{color}
\usepackage{txfonts}
\setcitestyle{numbers,super}
\begin{document}
\title{Hydrogen and Deuterium Tunneling in Niobium}
\date{\today}
\author{Abdulaziz Abogoda}
\affiliation{Hearne Institute for Theoretical Physics, Louisiana State University, Baton Rouge, LA 70803}
\affiliation{Applied Physics Graduate Program, Northwestern University, Evanston, IL 60208}
\author{W. A. Shelton}
\affiliation{Department of Physics and Astronomy, Louisiana State University, Baton Rouge, LA 70803}
\author{I. Vekhter}
\affiliation{Department of Physics and Astronomy, Louisiana State University, Baton Rouge, LA 70803}
\author{J. A. Sauls}\email{sauls@lsu.edu}
\affiliation{Hearne Institute for Theoretical Physics, Louisiana State University, Baton Rouge, LA 70803}

\begin{abstract}
We use density functional methods to identify the atomic configurations of H and D atoms trapped by O impurities embedded in bulk Nb.
The O atoms are located at the octahedral position in the Nb body-centered cubic (BCC) lattice, H (D) atoms tunnel between two degenerate tetrahedral sites separated by a mirror plane.
Using nudge elastic band (NEB) methods we calculate the double-well potential for O-H and O-D impurities, the wave functions and the tunnel splittings for H and D atoms.
Our results agree with those obtained from analysis of heat capacity and neutron scattering measurements on Nb with low concentrations of O-H and O-D.
\end{abstract}

\date{\today}

\maketitle

\vspace*{-1em}
\noindent{{\bf\emph{Introduction:}}
Niobium is central to a wide range of superconducting based technologies, from high Q superconducting radio frequency (SRF) cavities used for high-energy accelerators,~\cite{gra17} to resonators and circuits for applications in quantum computing, quantum sensing and detectors to search for dark matter and physics beyond the standard model, as well as gravitational wave detectors.~\cite{rom23,bog19,gao21,ber23,uek24}
Superconducting Nb films and resonators also provide a platform to study nonequilibrium superconductivity~\cite{zar23} and nonlinear microwave phenomena.~\cite{sau22}

Efforts to extend Nb's successes in accelerator technology to quantum device applications involving few photons at mK temperatures revealed microwave losses not present at higher temperatures or intense electromagnetic fields. The increased loss is seen in both 2D coplanar waveguide resonators as well as 3D SRF cavities, and is generally attributed to microwave absorption by two-level tunneling systems (TLSs).~\cite{sag11,rom17,mcr20}
Understanding the origin and effects of TLSs on Nb resonators and quantum circuits, and devising methods to mitigate TLS loss and decoherence, is a key area of research for developing superconducting based quantum processors and sensors.~\cite{dev13,mul19}

Many studies suggest that TLS losses result from surface oxides.~\cite{gao08,rom17,mul19,mcr20,rom20,alt22,baf24}
When Nb is exposed to air 5 to 7 nm of Nb oxide forms on the surface. The oxide consists mainly of disordered Nb$_2$O$_5$ on top of a monolayer of NbO$_{2-x}$.~\cite{ver21,she21}
As an example Fig.~\ref{fig-Q_vs_T} shows results reported in Ref.~\onlinecite{rom20} for the quality factor as a function temperature for Nb SRF cavities before and after heat treatment to dissolve and remove oxides from the Nb surface.~\footnote{The quality factor is the ratio of the photon lifetime, $T_1$, to the period, $2\pi/\omega$, i.e. $Q\equiv\omega T_1/2\pi$, and is a measure of the of efficiency of the SRF cavity in preserving the energy of the electromagnetic field.}
The decrease in $Q$ as $T$ drops to mK temperatures is characteristic of absorption from TLSs with tunnel splittings in the GHz range. Focusing on the 1.3 GHz Nb cavity the vertical black arrow shows the order of magnitude increase of Q after heat treatment to remove Nb oxides. A key result was that removal of Nb oxides leads to microwave photons with lifetimes of order $T_1\simeq 2\,\mbox{sec}$.~\cite{rom20}

Figure~\ref{fig-Q_vs_T} shows more: the orange arrow indicates that \emph{after} removal of Nb oxide there remains additional $T_1$ loss of nearly an order of magnitude that also shows the same characteristic of a distribution of TLSs.
The result suggests that TLSs may also be \emph{embedded} within the region of EM field penetration in Nb, i.e. the London penetration depth of order $\lambda_{L}\approx 50\,\mbox{nm}$. Indeed the heat treatment employed to dissolve Nb oxides will also lead to O diffusion into Nb. It is also established that H easily diffuses into Nb and can be trapped by O interstitials. These O-H (also N-H) interstitials in Nb were identified and studied in great detail starting in the late 1970s with inelastic neutron scattering, heat capacity and acoustic spectroscopy.\cite{pfe76,mor78,mag83,wip84,bel85,mag86}
A key study is the analysis of the low-temperature heat capacity of Nb infused with low concentrations ($x\simeq\{0.05\% - 1.4\%$) of oxygen-hydrogen (O-H) and oxygen-deuterium (O-D) by Wipf and Neumaier.~\cite{wip84} The authors adapted the theory of tunneling in amorphous solids~\cite{and72a,phi72} to show that H and D atoms undergo atomic tunneling between two interstitial sites in the Nb unit cell. 
Their analysis yielded a tunnel matrix element for O-H of $J_H = 0.190\,\mbox{meV}$ and $J_D = 0.021\,\mbox{meV}$ for O-D. The former agrees with the value for H based on inelastic neutron scattering.~\cite{wip81}

\begin{figure}
\includegraphics[width=\columnwidth]{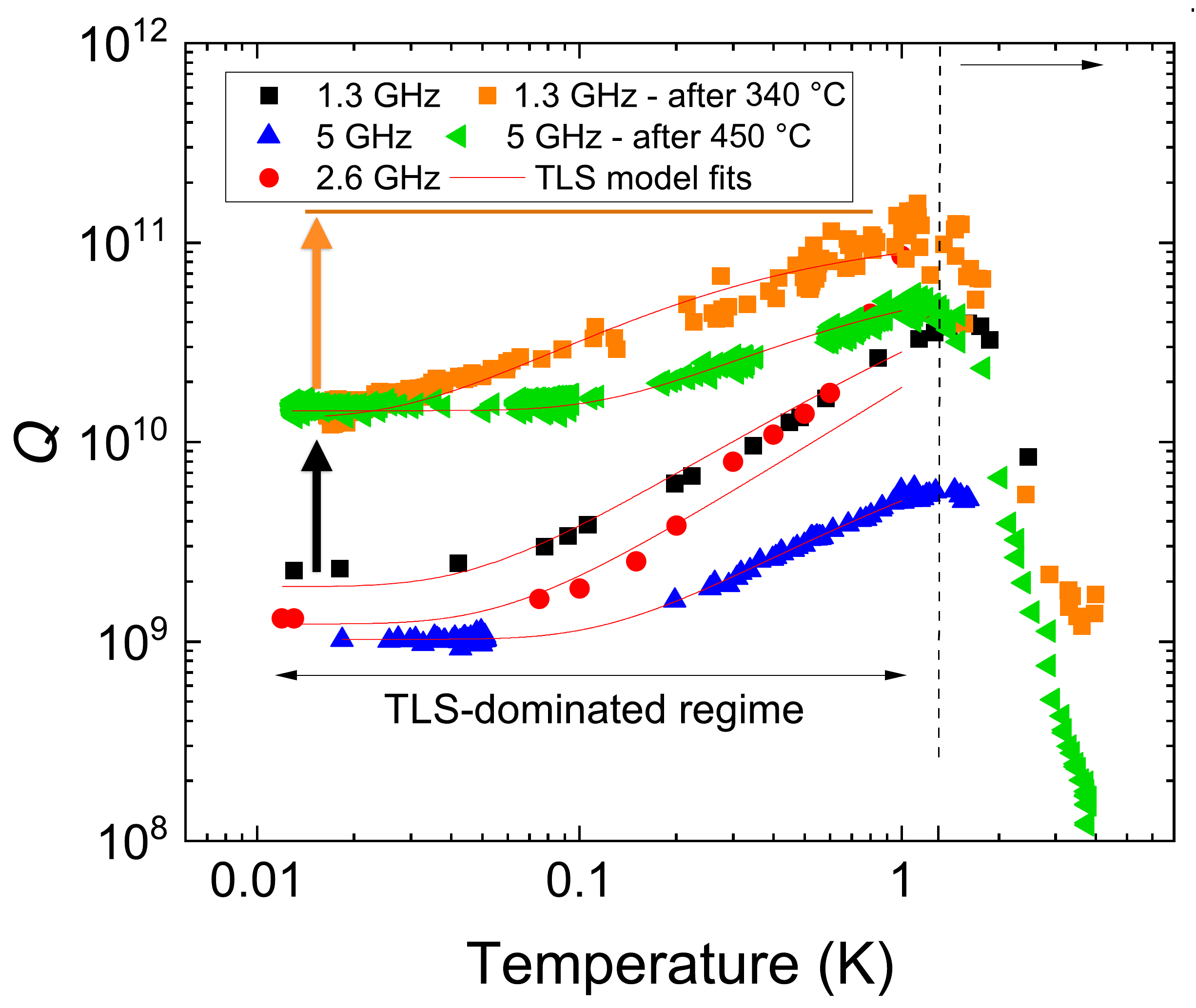}
\caption{Quality factor versus temperature for Nb SRF cavities at low power before and after removal Nb oxides by heat treatment. The black arrow shows the order of magnitude increase of Q for a 1.3 GHz cavity at the lowest temperature after removal of Nb oxides. The red arrow shows that even \emph{after removal} of Nb oxide residual $T_1$ loss of nearly an order of magnitude exists from a remaining distribution of TLSs. Adapted from Figure 4 of Ref.~\onlinecite{rom20}.}
\label{fig-Q_vs_T} 
\end{figure}

In this letter we report our results based on density functional theory (DFT) and related methods described below to compute the double-well potential for H trapped by O at interstitial sites within BCC Nb. 
We show that O is localized at the octahedral site and that there are several possible sites where H can be trapped in the near vicinity of O. These include the pair of tetrahedral sites proposed by Magerl et al.~\cite{mag83}
This configuration allows for quantum tunneling of H or D between two symmetry-equivalent positions separated by $1.26\,\mbox{\AA}$ and an energy barrier of $V_0 = 130\,\mbox{meV}$.
These parameters characterize the double-well potential for H or D which we obtain by computing the energy profile along the minimum energy path (MEP) as H is moved from one local minimum to the other.
The wavefunctions and energy levels of the ground and first excited state, and thus the tunnel splitting, are obtained by solving the stationary state Schr\"odinger equation for the MEP.
The calculated tunnel splittings for H and D compare well with the results based on low temperature heat capacity measurements and inelastic neutron scattering.~\cite{wip84} 
The results for the H and D tunnel splittings, both with and without effective mass corrections, are summarized in Table~\ref{table:tunnel_comp}. 
Note that D has tunnel splitting corresponding to $J_D/h=5.07$ GHz. Finite concentrations of OH and OD generate a random strain field and a distribution of tunnel splittings.~\cite{wip84} 
The natural abundance of D is $0.027$\%,~\cite{rum23} making embedded D TLSs a potential source  of microwave loss for Nb devices operating the GHz frequency range.

\vspace*{-1em}
\medskip\noindent{{\bf\emph{Methods:}}
We used the Quantum Espresso (QE) package to perform DFT calculations.\cite{gia09,gia17,gia20} The QE Plane-Wave Self-Consistent Field (PW) module was used for the initial ionic minimizations. Supercells of BCC Nb of dimension $3\times 3\times 3$ lattice spacings were used to minimize interactions between periodic images of the interstitials.
We identify the most probable tunneling path for H or D as the MEP. The DFT calculations for H are applicable to deuterium (D) because the mass enters only the dynamics of the tunneling process, not the potential energy curve traced by the MEP.
Two approaches were used to calculate the MEP: Nudge Elastic Band (NEB), and Variable Cell Nudge Elastic Band (VCNEB). The NEB calculation was performed using QE's NEB module.
The VCNEB calculation was performed using VCNEB module provided by Universal Structure Predictor: Evolutionary Xtallography software package (USPEX) while using QE as the DFT engine.~\cite{qia13} 
Both methods calculate the MEP by simulating intermediate ``images'' of the system between the initial and final configurations. Each atom is connected to their respective copies in the adjoining images via virtual springs to constrain them to remain on a continuous path.~\cite{jon98} Minimizing the total force on the atoms in each image results in the MEP. The VCNEB method differs from the NEB that it allows the cell (axes, angles) to also vary along the path. 
From the MEP we construct the potential energy as a function of a reaction coordinate that measures the distance traveled by the all atoms in the supercell along the MEP. 
The results for the potential barrier for both methods are discussed in~\autoref{sec-HD_tunneling}\emph{H and D Tunneling}.

The DFT parameters were chosen to ensure that the energy per atom in every calculation converges within $1.36$ meV. Relaxations of the ions were set with the same threshold and involve moving all cell axes and angles unless otherwise specified.
The DFT self-consistency threshold was set to $1.36\times 10^{-8}$ eV per atom to ensure accurate force calculations. An energy cutoff for the plane wave basis of $1088$ eV was used. The calculations were performed over a uniform Monkhorst-Pack grid of k-points with a resolution of 
$0.1$\AA$^{-1}$, and utilized Marzari-Vanderbilt smearing with a value of $0.054$ eV. These parameters were found by performing convergence tests relative to energy/atom on unit cells of Nb, NbO, and NbH, and then confirmed with Nb supercells with H and O interstitials.
Perdew–Burke–Ernzerhof (PBE) exchange-correlation functionals were used. The Oxygen and Niobium pseudopotentials were Projector Augmented Wave (PAW) while the Hydrogen used an Optimized Norm-Conservinng Vanderbilt pseudopotential (ONCVPSP).
With these parameters the Nb lattice parameter converges to $3.3$\AA. Force and pressure convergence thresholds for NEB and VCNEB calculations were determined by convergence testing relative to the energy/atom standard of $1.36$ meV.
A summary of the parameters are listed in Table~\ref{table-DFT_parameters}, while the exchange-correlation functionals are from Ref.~\onlinecite{pra23}.

\begin{table}[t]
\begin{tabular}{|c|c|c|c|c|c|c|}
\hline
method & 
$V_0$[meV] & 
d[\AA] & 
$J_{H}$[meV] & 
$J_{D}$[meV] & 
$m_{H}^*$ & 
$m_{D}^*$ \\
\hline
{\footnotesize NEB}   & 134 & 1.25 & 0.507 & 0.0334 & 1.32 & 1.10 \\
\hline
{\footnotesize VCNEB} & 130 & 1.26 & 0.358 & 0.0187 & 1.18 & 0.977\\
\hline
\multicolumn{3}{|l|}{Heat capacity measurement~\cite{wip84}} & 0.19 & 0.021 & \multicolumn{2}{l|}{ }\\
\hline
\end{tabular}
\caption{Tunnel barrier, path length, and tunnel splitting for H and D for both 
barrier potentials obtained from both NEB and VCNEB.
The calculated tunnel splittings of H and D assume their bare masses, $m_{H,D}$.
The effective masses, $m^*_{H,D}$ [$m_{H,D}$], are those that precisely reproduce the 
experimental tunnel splittings.}
\label{table:tunnel_comp} 
\end{table}

\begin{figure}[H]
\includegraphics[width=\columnwidth]{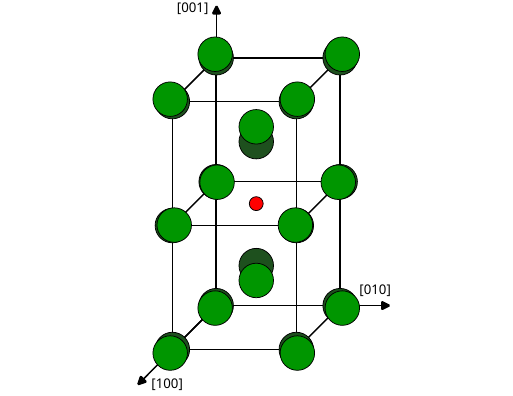}
\caption{
The strained BCC lattice with Nb atoms shown in green. The O atom shown in red is located at the octahedral position. Unstrained positions of Nb atoms are shown in dark green to highlight the deformation of the lattice by the O impurity.
The two unit cells shown were extracted from the $3\times 3\times 3$ supercell.}
\label{fig:NbO}
\end{figure}

\vspace*{-1em}
\medskip\noindent{{\bf\emph{O and H Interstials:}}
We first examine the effects of O or H intersitials on the $3\times3\times3$ supercell, and identify energetically stable positions for both atoms. We find that O is stable only in the octahedral site.
The O atom increases the volume of the $3\times 3\times 3$ supercell by 17{\AA}$^3$ ($+1.7\%$), and the distance between the O atom and the nearest Nb atom is 2.0{\AA}. This is significant lattice strain as can be seen in Fig.~\ref{fig:NbO}.

By itself H is stable in both the octahedral and tetrahedral positions of the unit cell and increases the volume by 12\AA$^3$ ($+1.2\%$) for both configurations shown in Fig.~\ref{fig:NbH}.
The tetrahedral configuration has the lower energy and shows slightly less local strain of the Nb atoms in the supercell.
X-ray scattering measurements report a volume expansion of 3.13\,\AA$^3$ per H atom.~\cite{pfe77} We do not have an explanation for the discrepancy with our result.
However, the distance from H to the nearest Nb atom is calculated to be $1.93\,{\AA}$ for H in the tetrahedral configuration.
This compares with $1.94\,\AA$ reported in Ref.~\onlinecite{blo10}. Our result corresponds to a strain displacement of the nearest Nb of $0.09\,\AA$. This value compares well with the value of $0.08\,\AA$ reported in Ref.~\onlinecite{sun04} for H trapped at a tetrahedral site.
The O and H sites proposed by Magerl et al.~\cite{mag83} (``c sites'') for H tunneling are shown in Fig.~\ref{fig:NbOH}(a). The O atom located at the octahedral site captures an H atom at a tetrahedral site labelled ``c''.  There are two such ``c'' sites separated by a mirror plane and an energy barrier through which the H atom can undergo quantum tunneling and form a TLS impurity.

\begin{figure}[t]
\includegraphics[width=\columnwidth]{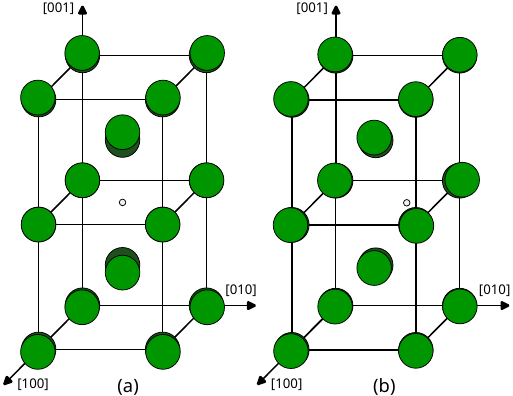}
\caption{
The strained BCC lattice with Nb atoms shown in green. The H atom shown in gray is located at the octahedral [tetrahedral] position in panel (a) [(b)]. Unstrained positions of Nb atoms are shown in dark green to highlight the deformation of the lattice by the O impurity.
The two unit cells shown were extracted from the 
$3\times 3\times 3$ 
supercell.
}
\label{fig:NbH} 
\end{figure}

The presence of strain in the Nb crystal due of a finite concentration of O-H impurities or other defects introduces an asymmetry energy, $\epsilon$, between the two local minima of the double-well potential, and thus for any one O-H impurity an energy splitting between the ground and excited state of $E=\sqrt{J^2+\epsilon^2}$.~\cite{wip84} Here we calculate the tunneling splitting of an \emph{isolated} O-H or O-D impurity in an otherwise BCC Nb crystal. We address the role of strain induced by a finite concentration of impurities, and the resulting distribution of tunnel splittings in a separate report.
 
\phantomsection
\label{sec-HD_tunneling}
\vspace*{-1em}
\medskip\noindent{{\bf\emph{H and D Tunneling:}}
In the standard NEB method the cell dimensions of the initial and final configurations as well as the intermediate configurations are equal. However, the optimally relaxed configuration described in the previous section has a cell that is a non-rectangular cuboid with different $x$ and $y$ dimensions. Furthermore, the relaxed starting and ending configurations of the MEP between the two ``c'' positions do not have the same cell dimensions.
This suggests we allow adjustable cell dimensions that also enforce mirror symmetry across the $x=y$ plane. Thus, we allow the cell to relax in the $z$ direction while keeping the $x$ and $y$ dimensions constant and equal to the average of the $x$ and $y$ dimensions of the optimal supercell, thus enforcing a rectangular cell that respects the mirror symmetry across the $x=y$ plane. We also make use of this symmetry to compute only half of the MEP and potential energy profile. 
First we find the tunnel barrier maximum at the midpoint of the MEP using the Climbing Image (CI) NEB technique,\cite{jon98,hen00} then calculate the MEP from the initial configuration to the midpoint using NEB with seven intermediate images.~\cite{jon98} The full path is obtained by mirror symmetry. We then end up with 17 points defining the MEP, including the two endpoints.

\begin{figure}[t]
\includegraphics[width=\columnwidth]{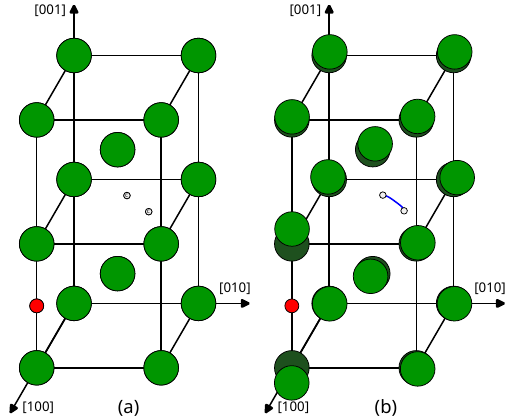}
\caption{
Panel (a): The configuration showing O at the octahedral site and two equivalent tetrahedral sites (``c'' sites) for H in the un-relaxed BCC lattice.
Panel (b): The relaxed configuration showing the strain induced by both O and H impurities. The MEP of the H atom is shown in blue.
}
\label{fig:NbOH} 
\end{figure}

\begin{figure}[h]
\includegraphics[width=1\columnwidth]{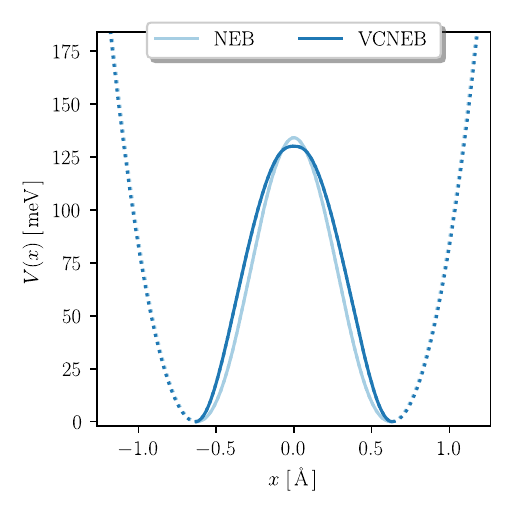}
\caption{The double-well potential energy $V(x)$ based on the energy profile of the MEP calculated using the NEB and VCNEB methods as a function of the reaction coordinate $x$. The smooth curves are cubic splines fit to the values calculated at the VCNEB points. The dashed curves are half harmonic potentials for the region exterior to the MEP.}
\label{fig:barrier} 
\end{figure}

The VCNEB method while computationally more expensive does not require us to enforce special constraints on the cell axes, as the method allows the cell to transform along the MEP. Similar to our NEB computation, the midpoint is first obtained using CI technique and then half of the path is calculated from one endpoint to the midpoint with 5 intermediate images.

The potential energy profiles, $V_{\mbox{\tiny MEP}}(x)$, obtained using the NEB and VCNEB methods are shown as the solid lines in Fig.~\ref{fig:barrier}. The NEB potential has a maximum barrier height of 134 meV and a reaction path length of 1.25\AA. The H atom travels 1.142\AA between the two relaxed ``c'' sites. The difference between these two path lengths is indicative of the total displacement of the rest of the atoms in the supercell.
Note that VCNEB potential energy profile and barrier height does not differ much from 
those obtained using the NEB method, $V_0=130\,\mbox{meV}$ and a reaction path length of 1.26\AA.
Nevertheless, these differences modify the tunnel splittings for O-H and O-D TLS impurities as we discuss below.

The full double-well potential for H tunneling is constructed by adding a half harmonic potential of the form $V_{\mbox{\tiny h $\pm$}}(x)=\nicefrac{1}{2}\,K\,(x\pm d/2)^2$ to the regions bordering the MEP.
The elastic constant $K$ of the half harmonic wells is calculated by moving the H around its relaxed position along the 6 principal directions in small increments, $\Delta x$, relaxing the supercell around it to compute $\Delta V(x)/\Delta x$  using QE. The resulting half harmonic wells are shown as the dashed lines in Fig.~\ref{fig:barrier}, with the double-well potential given by $V(x)=V_{\mbox{\tiny MEP}}(x)+V_{\mbox{\tiny h $-$}}(x)+V_{\mbox{\tiny h $+$}}(x)$.

To calculate the tunnel splitting we solve the Schr\"odinger equation, 
$-\frac{\hbar^2}{2m^*}\,\phi_{xx}(x) + V(x)\,\phi(x) = E\,\phi(x)$
using the NumPy eigensolver defined on an equally spaced discretized reaction path,
for the two lowest energy eigenstates, $E_{g,e}$, and their corresponding wave functions, 
$\phi_{g,e}(x)$, where $m^*$ is the effective mass of the tunneling atom. 
Assuming the bare mass of H, the tunnel splitting, $J=E_e-E_g$, based on the NEB barrier is calculated to be $J_H = 0.507$ meV, while for the bare mass of deuterium we obtain $J_D=0.0334$ meV.
For comparison Wipf and Neumaier report values of $J_H = 0.19$ meV and $J_D=0.021$ meV from analysis of their heat capacity measurements based on a variation of the tunneling theory for amorphous solids,~\cite{wip84} which is reasonable agreement given the sensitivity of the tunnel splitting to the barrier height and width.  
That sensitivity is apparent when we calculate the tunnel splittings based on the VCNEB barrier potential: $J_H=0.358$ meV and $J_D = 0.0187$ meV, corresponding 
to a 29\% (44\%) reduction in the H (D) tunnel splitting compared to those based 
on the NEB barrier potential, even though the barrier height is reduced by only 3\%.
Note also the closer agreement with the experimental values of Wipf and Neumaier, particularly for deuterium.
The energy levels and wavefunctions computed from the VCNEB barrier potential are also shown in Fig.~\ref{fig:schrodinger}.

\begin{figure}
\includegraphics[width=1\columnwidth]{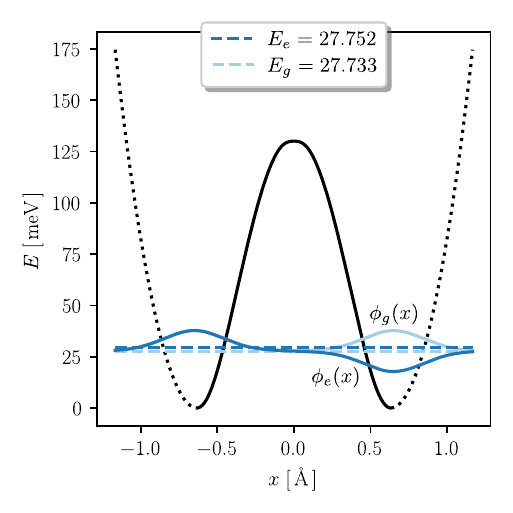}
\caption{The double-well potential $V(x)$ based on the VCNEB potential barrier (solid black) combined with the half-harmonic potentials (dotted black) obtained from DFT as a function of the reaction coordinate $x$. The dashed lines indicate the two lowest energy levels with values in meV indicated in the legend. The solid lines are the corresponding ground (light blue) and excited (dark blue) state wave-functions.}
\label{fig:schrodinger} 
\end{figure}	

The difference in the length of the reaction path compared to the path length traveled by H or D is indicative of the interactions of tunneling atom with the nearby atoms in the supercell. These interactions give rise to an effective mass of the tunneling atom. Given that the difference in reaction path and the path length traversed by H is $\approx 0.12\,\mbox{\AA}$ compared to $1.14\,\mbox{\AA}$ we do not expect large corrections to the bare mass of the tunneling atoms.

If the tunnel splittings obtained from analysis of the heat capacity of H and D tunneling in Nb are accurate values for \emph{isolated} O-H and O-D impurities then we can determine the effective masses by requiring the splitting between energy eigenvalues obtained from the Schr\"odinger equation to agree with the experimentally determined values.
This analysis results in effective masses of $1.32 m_{H}$ and $ 1.10 m_{D}$ for the MEP computed using the NEB method, and $1.18 m_{H}$ and $0.977 m_{D}$ for the MEP based on the VCNEB methods, indeed small corrections to the bare masses of H and D.
Table (\ref{table:tunnel_comp}) summarizes the tunnel barrier heights, reaction path lengths, tunnel splittings and effective masses we obtain from both the NEB and VCNEB methods.

It is worth commenting on earlier theoretical reports of H(D) tunneling in pure Nb, i.e. H(D) tunneling in the absence of trapping by O or N. Sundell and Wahnström~\cite{sun04} reported results based on DFT for the potential energy profile and level splittings for H(D) tunneling between tetrahedral c-sites - $V_0\simeq 200$ meV and $J_H=0.8$ meV ($J_D=0.1$ meV) - that differ significantly from our results for H(D) tunneling with O at the octahedral site, as well as the results obtained from analysis of heat capacity and inelastic scattering data for dilute concentrations of O-H and O-D as summarized in Table~\ref{table:tunnel_comp}.
The differences likely reflect the sensitivity of the tunnel splittings to the potential energy profile, which differs due to trapping of H(D) by O, as well as the deformation of the unit cell by O. 
These results also reflect differences in the method used to determine the potential energy profile and tunneling barrier. 
The VCNEB method we use was developed to determine accurate atomic positions of all atoms for the MEP.
Earlier calculations by Schober and Stoneham reported a much smaller tunnel splitting of $J_H=0.01$ meV.~\cite{sch88} However, this result is based on a phenomenological Nb-H potential rather than DFT methods to determine the potential energy profile.
Although outside the scope of this Letter, the Nb-H(D) system deserves further study using the methodology we developed for NbOH(D).

\medskip\noindent{{\bf\emph{Summary:}}
We have demonstrated that the configuration of H trapped by O in BCC Nb as proposed by 
Magerl et al.,~\cite{mag83} i.e. O atoms localized at the octahedral site and H atoms tunneling between the tetrahedral ``c'' sites in the unit cell, is indeed plausible based on (i) the stability of O at only the octahedral site, (ii) stability of H at the tetrahedral ``c'' sites and (iii) good agreement between the calculated tunnel splittings for O-H and O-D and the values inferred from heat capacity measurements and inelastic neutron scattering.
The reasonable values for the effective masses inferred from exact match the experimental tunnel splittings to the calculated values also supports this conclusion.

\medskip\noindent{{\bf\emph{Outlook:}}
We tested all possible unique positions for trapping H atoms within one unit cell of the octahedral O. Several stable positions of H have a lower energy than the tetrahedral ``c'' positions shown in Fig.~\ref{fig:NbOH}. Three such configurations also have an equivalent mirror site that may allow for tunneling through a symmetric double well potential similar to the Magerl configuration. Thus, there may be a spectrum of tunnel splittings that so far have not been studied experimentally or theoretically. Depending on the potential barrier and reaction path there may be embedded O-H TLSs that are directly relevant to decoherence and loss in Nb. These configurations warrant further study.
For similar reasons the role of strain generated from a finite concentration of O-H and O-D impurities, as well as structural defects, on the distribution to tunnel splittings and asymmetry energies also warrants further study using ab initio DFT methods.

\eject
\noindent{{\bf\emph{Acknowledgements:}}
We thank Alex Romanenko and Anna Grassellino at Fermilab for discussions on materials issues limiting the quality factor in microwave resonators that led to this study.
The research of AA and JAS was supported by the U.S. Department of Energy, Office of Science, National Quantum Information Science Research Centers, Superconducting Quantum Materials and Systems Center (SQMS) under contract number DE-AC02-07CH11359.
WAS and IV are supported in part by the US Air Force Sponsored Office of Research award FA9550-23-1-0709.

\onecolumngrid
\appendix
\section{DFT Parameters and Nb displacements}

\vspace*{-5mm}
\begin{table}[H]
\centering
\begin{tabular}{|ll|}
\hline
Parameter & Value \\
\hline
convergence threshold for choosing DFT parameters & $1\times 10^{-4}$ Ry = $1.36$ meV \\
\hline
convergence threshold on energy/atom for DFT self consistency & $1\times 10^{-9}$ Ry = $1.36\times 10^{-8}$ eV \\
\hline
convergence threshold on energy/atom for BFGS ionic minimization & $1\times 10^{-4}$ Ry = $1.36$ meV \\
\hline
charge density kinetic energy cutoff & $80$ Ry = $1088$ eV \\
\hline
smearing factor & $0.004$ Ry = $0.054$ eV \\
\hline
k-points sampling (distance between k-points) & $0.1$\AA$^{-1}$ e.g. 
$6\times 6\times 6$ 
for a 
$3\times 3\times 3$ 
Nb supercell \\
\hline
\end{tabular}
\caption{The DFT parameters used in this study.}
\label{table-DFT_parameters}
\end{table}

\begin{figure}[h]
\begin{minipage}[c]{0.9\textwidth}
\includegraphics[width=0.6\textwidth]{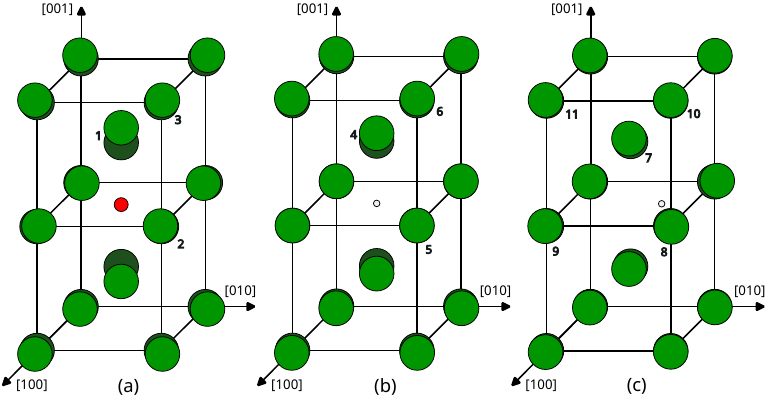}
\caption{Nb displacements for single O and H interstital impurities. Panel (a): O at the octahedral site. Panel (b): H at the octahedral site. Panel (c): H at the tetrahedral site. The 11 unique displacements are indexed and quantified in Table~\ref{table:deformation}.}
\label{fig:7}
\end{minipage}
\end{figure}
\begin{table}[h]
\begin{minipage}[c]{0.8\textwidth}
\begin{tabular}{|c|c|c|c|c|}
	\hline
	Panel & Atom & [100] & [010] & [001] \\
	\hline
	a & 1 & 0.000 & 0.000 & 0.395 \\
	\hline
	a & 2 & -0.037 & -0.037 & 0.000 \\
	\hline
	a & 3 & 0.043 & 0.043 & 0.079 \\
	\hline
	b & 4 & 0.000 & 0.000 & 0.203 \\
	\hline
	b & 5 & 0.003 & 0.003 & 0.000 \\
	\hline
	b & 6 & 0.015 & 0.015 & 0.057 \\
	\hline
	c & 7 & 0.000 & -0.043 & 0.072 \\
	\hline
	c & 8 & 0.072 & 0.043 & 0.000 \\
	\hline
	c & 9 & 0.018 & -0.030 & 0.000 \\
	\hline
	c & 10 & 0.008 & 0.003 & 0.022 \\
	\hline
	c & 11 & 0.013 & -0.020 & 0.028 \\
	\hline
\end{tabular}
\caption{Displacements of Nb atoms [in \AA] generated by O and H interstitials as 
         indexed in Fig.~\ref{fig:7}.
\label{table:deformation} 
}
\end{minipage}
\end{table}
\vspace*{-10mm}
%
\end{document}